# Factors Determining the Carrier Distribution in InGaN/GaN Multiple-Quantum-Well Light-Emitting Diodes

Dong-Pyo Han[1, 2 a)], Jong-In Shim[2], and Dong-Soo Shin[3]

[1]*Faculty of Science and Technology, Meijo University, 1-501 Shiogamaguchi, Tempaku-ku, Nagoya 468-8502, Japan*

[2]*Dept. of Electronics and Communication Engineering, Hanyang University, ERICA Campus, Ansan, Gyeonggi-do 426-791, Korea*

[3]*Dept. of Applied Physics and Dept. of Bionanotechnology, Hanyang University, ERICA Campus, Ansan, Gyeonggi-do 426-791, Korea*

**Abstract**

Factors determining the carrier distribution in InGaN/GaN multiple-quantum-well (MQW) light-emitting diodes (LEDs) are studied via photoluminescence and temperature-dependent electroluminescence spectra. Employing a dichromatic LED device, we demonstrate that the carrier recombination rate should be considered playing an important role in determining the carrier distribution in the MQW active region, not just the simple hole characteristics such as low mobility and large effective mass.

[a)] E-mail: han@meiju-u.ac.jp

# 1. Introduction

In the last decade, significant progress has been made in nitride-based semiconductor devices, especially the blue light-emitting diodes (LEDs) with the InGaN/GaN multiple-quantum-well (MQW) structure as the active layer [1],[2]. Based on this progress, high-brightness GaN-based LEDs are replacing the conventional lighting such as incandescent light bulbs and fluorescent lamps for their low power consumption and high energy-conversion efficiency. However, many issues still exist with regard to carrier recombination, transport, and distribution in the MQWs utilized in blue LED devices. Consequently, the devices suffer from reductions in external quantum efficiency (EQE) at high current densities. Still, achieving high efficiency at high current densities is required to expedite the use of LEDs as general lighting sources [3],[4].

Recently, a reduced effective active volume was reported to be induced by nonuniform carrier distributions in the MQWs, which could have a significant influence on the device performance such as the EQE and the applied voltage at high current densities [3]-[6]. It is typically believed that the inhomogeneous carrier distribution in MQWs originates from differences in mobility, effective mass, and carrier concentration between electrons and holes in InGaN-based LED devices. The mobility of holes is only a few $cm^2/V \cdot s$, whereas that of electrons is several hundred $cm^2/V \cdot s$. Moreover, the effective mass and the activation energy are known to be $\sim 1.1 m_0$ and $\sim 150$ meV for holes and $\sim 0.2 m_0$ and $\sim 20$ meV for electrons, respectively [7],[8]. As a consequence, holes are typically known to be distributed predominantly in the quantum wells (QWs) closest to the p-type layer. From this reason, hole injection is thought to determine the carrier distribution in the MQW active region and thus the effective active volume in GaN-based LEDs [9].

In this paper, in order to clarify the carrier distribution in InGaN/GaN MQW LEDs and its determining factors, dual-wavelength (or dichromatic) MQWs are employed. While there have been researches using dichromatic or trichromatic MQW LEDs with different indium compositions in each QW [10],[11], factors determining the carrier distribution in InGaN/GaN MQW LEDs have not yet been clearly elucidated. Dichromatic or trichromatic LEDs are useful to investigate the distribution of carriers in MQWs as emission wavelengths can be used to determine the carrier distribution. Since it is imperative to have a uniform distribution of carriers across all QWs in order to realize highly efficient LED devices, the study of factors determining the carrier distribution in MQW LEDs is of prime importance.

## 2. Experiments

Figures 1 (a) and (b) schematically depict the epitaxial structure and the MQW active region of the sample used in this study. The sample has been grown by metal-organic chemical vapor deposition (MOCVD) on a *c*-plane sapphire substrate. The structure is composed of, from the substrate, a low-temperature nucleation GaN layer, a Si-doped n-GaN layer (n-doping = $5\times10^{18}$ cm$^{-3}$), and the MQWs consisting of seven pairs of 3.5-nm-thick QWs sandwiched by 5-nm-thick GaN barriers. The top and bottom QWs have an indium composition of 12% while the other five QWs in the center possess an indium composition of 15%. On top of the MQWs are a 20-nm-thick p-AlGaN electron-blocking layer (EBL) and a Mg-doped p-GaN layer (p-doping = $1\times10^{18}$ cm$^{-3}$). The sample has lateral electrodes with a chip size of 700×270 m$^2$, mounted on a surface-mount-device (SMD) package.
For experiments, we utilize photoluminescence (PL) and electroluminescence (EL) spectra from the dichromatic LED device and analyze their emission characteristics. In particular, EL spectra are meticulously investigated for low operating temperatures from 50 to 200 K, with consideration of carrier recombination rates in MQWs. From the analysis, we attempt to understand the underlying factors determining the carrier distribution in MQWs.

## 3. Measurement results and discussion

Figures 2 (a) and (b) show the PL spectra of the sample at room temperature, plotted on linear and log scales, respectively. A 325-nm He-Cd laser has been used as an external pumping source and spectra are normalized to the peak intensity. As illustrated in Fig. 2 (a), two emission peaks are observed, i.e., one at ~445 nm (spectrum 1) and the other at ~430 nm (spectrum 2). Spectrum 1 and spectrum 2 correspond to the recombination of electron-hole (e-h) pairs in In$_{0.15}$Ga$_{0.85}$N and In$_{0.12}$Ga$_{0.88}$N QWs, respectively. The dashed lines in Fig. 2 (a) are fitting curves of the PL spectrum with Gaussian functions [12]. A discrepancy between the fitting curve and the experimental PL spectrum is noted in the long-wavelength region (460 - 480 nm). Since this long-wavelength shoulder does not affect our subsequent analysis, it is ignored afterwards.

The ratio of the integrated Gaussian fitting for spectrum 1 to the total integrated spectrum is ~0.75, which is similar to the ratio of the number of QWs corresponding to spectrum 1 to the total number of QWs (5/7 = ~0.71). In general, almost an equal number of e-h pairs are generated and recombine at each QW in PL measurements [8]. With increase in intensity of the external pumping source, the peak wavelengths of both spectra show slight blue shifts, but the ratio between two integrated spectra still remains at similar values. Thus, we can infer that if the carriers are uniformly distributed across all QWs under forward bias, the EL spectrum should be similar to the PL spectrum shown in Fig. 2, with spectrum 1 more pronounced than spectrum 2.

Figure 3 presents the EL spectra dependent on the driving current at room temperature. Each spectrum is normalized to its peak value. In the low-driving-current region (< 10 mA), it is observed that spectrum 1 emitted from the lower-bandgap $In_{0.15}Ga_{0.85}N$ QWs is predominant. However, as the driving current is increased, spectrum 2 emitted from the higher-bandgap $In_{0.12}Ga_{0.88}N$ QWs increases drastically compared to spectrum 1. In the high-driving-current region (> 150 mA), spectrum 2 becomes more pronounced and the peak wavelength of the EL spectrum shifts from ~445 nm towards ~430 nm. From the experimental results shown in Fig. 3, we notice that the recombination of e-h pairs initially occurs preferentially at the $In_{0.15}Ga_{0.85}N$ QWs, which are located in the central region of the MQWs. Then, as the current increases, the recombination of e-h pairs in the QWs closest to the p- or n-type layers becomes more significant. The experimental results in Fig. 3 are contrasted with the conventional understanding of the carrier distribution in the MQWs, i.e., holes predominantly occupy the QWs closest to the p-type layer due to their high effective mass and low mobility compared to those of electrons, resulting in preferential radiative recombination there [13],[14]. Obviously, the QWs with different bandgaps have affected the carrier distribution. The fact that the lower-bandgap QWs emit first at lower currents is considered to be caused by the carrier concentrations building up there more rapidly than in the higher-bandgap QWs, thus inducing more e-h recombinations. These experimental results seem to indicate that different characteristics of holes and electrons are not the only dominant factors determining the carrier distribution in the MQWs. Rather, it may be the carrier dynamics that determines the eventual carrier distribution in the MQWs. The observation that at high driving currents, spectrum 2 from the higher-bandgap QWs occupies a more significant portion than in the PL spectrum shown in Fig. 2 is discussed in detail later with numerical simulations.

To deepen our understanding, we have investigated the EL spectra of our sample from 50 to 200 K. A helium closed-cycle cryostat from Advanced Research Systems has been used to cool the device to cryogenic temperatures with thermal grease applied to ensure good thermal conductivity. Figures 4 (a) and (b) show the EL spectra dependent on the operating temperature at driving currents of 10 μA and 10 mA, respectively. Again, each spectrum is normalized to its peak value. The EL spectra can be clearly differentiated between spectrum 1 and spectrum 2 due to the nature of decreases in the full width at half maximum (FWHM) of the EL spectra at low operating temperatures [15]. As depicted in Fig. 4 (a), at a low driving current of 10 μA, the primary emission spectrum is spectrum 1 and the ratio of spectrum 2 to spectrum 1 increases gradually as the operating temperature decreases. On the other hand, in Fig. 4 (b) measured at a high driving current of 10 mA, the EL spectrum shifts almost completely from spectrum 1 to spectrum 2 and the ratio of spectrum 2 to spectrum 1 increases dramatically as the temperature decreases. In summary, we observe the overall behavior that the portion of spectrum 2 increases compared to spectrum 1 as the operating temperature decreases.

For a more detailed analysis of experimental results in Figs. 4 (a) and (b), the integrated powers of spectrum 1 and spectrum 2 are plotted as a function of driving current at temperatures of 50 and 200 K in Figs. 5 (a) and (b), respectively. It is observed that the rate of increase (slope) of spectrum 1 becomes smaller after a certain driving current whereas spectrum 2 begins at a lower value than spectrum 1, catches up with spectrum 1, and eventually follows the behavior of spectrum 1. The only difference between the phenomena observed in Figs. 5 (a) and (b) is the driving current at which the slope becomes smaller, which is roughly equal to the crossing point of spectrum 1 and spectrum 2. In all cases, QWs with the lower bandgap energy near the center of the MQWs emit predominantly at low currents even though they are not the ones closest to the p-type layer. At high currents, emissions from QWs close to the p- and n-type layers possessing a higher bandgap energy becomes more pronounced.

The reason why the overall EL efficiency at 50 K is lower and starts showing a lower slope at a lower current than at 200 K is that at 50 K, the total recombination rate in the active region becomes saturated at a lower current than at 200 K. The total recombination rate in the active region of an InGaN-based LED device is significantly affected by the operating temperature [16], [17]. At cryogenic temperatures, the Shockley-Read-Hall (SRH) recombination via defect levels, which is a primary nonradiative recombination process inside the MQWs, is suppressed as defects become inactivated. Consequently, the total recombination rate in MQWs decreases as the operating temperature decreases. As the carrier injection rate exceeds the total carrier recombination rate at the lower-bandgap $In_{0.15}Ga_{0.85}N$ QWs, carriers start to spill over to adjacent higher-bandgap $In_{0.12}Ga_{0.88}N$ QWs and more light is emitted from there.

A similar argument can be made as to why spectrum 2 becomes more pronounced as the temperature is lowered to 50 K at a given current level [Figs. 4 (a) and (b)]. In this case, the total recombination rate in the lower-bandgap QWs becomes smaller as the temperature is lowered due to the same reason mentioned above. Consequently, the carriers spill over to the adjacent higher-bandgap QWs, enhancing the emission there. From this reasoning, we can infer that the recombination rate in the QWs plays an important and decisive role in determining the carrier distribution in MQWs as much as the hole characteristics.

In order to investigate the carrier distribution and its determining factor from a different point of view, numerical simulations have been performed at 300 K by using the APSYS software. Most of the parameters used in this work are the same as in [18] except the QW structure and doping concentrations. Figures 6 (a), (b), and (c) depict the two-dimensional distributions of electrons and holes, and the radiative recombination rate in the active region, respectively. Cases of two different current densities are shown: low current density (3.5 A/cm$^2$) and high current density (70 A/cm$^2$). Each data is normalized to its peak value for comparison. The bimolecular radiative recombination coefficient ($B$) and the Auger recombination coefficient ($C$) are intentionally changed from $5 \times 10^{-18}$ cm$^3$/s and $5 \times 10^{-32}$ cm$^6$/s ($1 \times 10^{-17}$ cm$^3$/s and $1 \times 10^{-33}$ cm$^6$/s) at the low current density to $1 \times 10^{-18}$ cm$^3$/s and $1 \times 10^{-32}$ cm$^6$/s ($5 \times 10^{-18}$ cm$^3$/s and $5 \times 10^{-34}$ cm$^6$/s) at high current density for $In_{0.15}Ga_{0.85}N$ ($In_{0.12}Ga_{0.88}N$) QWs, respectively, to include the phase-space filling effect in these numerical simulations. [19]-[20] It is noted that without the change in $B$ and $C$ coefficients, the change in spectrum could not be reproduced. At the low current density, electrons mostly occupy higher-bandgap QWs on the n- and p-sides, whereas holes are mostly accumulated in the lower-bandgap QWs in the central region. As a result, the radiative recombination process can be observed mainly from the QWs in the central region, contributing to spectrum 1, since the scarce hole distribution is the dominant factor in determining the recombination of e-h pairs. On the other hand, at the high current density, the electron density in the higher-bandgap QWs slightly increases compared to the one at the low current density. Meanwhile, the hole density in higher-bandgap QWs on n- and p-side drastically increases compared to the one at the low current density. As a result, the radiative recombination process can now be observed dominantly from the higher-bandgap QWs, contributing to spectrum 2. This characteristics of nonuniform carrier distributions depending on the injection current level explains the difference between the spectra from the EL and the uniformly pumped PL and also highlight the importance of the recombination rate and associated energy levels in QWs in the active region in the LED under current injection.

In Figs. 7 (a) and (b), we show schematic energy band diagrams, including carrier injection, recombination, and distribution at low and high driving currents, for summary and discussion of what has been explained. At low driving currents (forward bias) [Fig. 7 (a)], injected holes from the bulk layers begin to occupy the lowest energy states in the active region within the diffusion length of hole and the carrier concentrations build up there while the electrons occupy all the QWs. This results in spectrum 1 being emitted via recombination of e-h pairs in $In_{0.15}Ga_{0.85}N$ QWs. As the driving current (forward bias) is increased further, quasi-fermi level approach the $In_{0.12}Ga_{0.88}N$ QW consequently hole begin to occupy QW closest to the p-type layer. Here the QW closest to the n-type layer is thought to be out of diffusion length of hole. Simultaneously, the carriers fill up quantized energy states in the $In_{0.15}Ga_{0.85}N$ QWs consecutively due to the carrier injection rate exceeding the recombination rate, a phenomenon known as the phase space filling and band filling effect [6],[15]. Eventually, beyond a certain carrier concentration in the $In_{0.15}Ga_{0.85}N$ QWs, more carriers spill over to the $In_{0.12}Ga_{0.88}N$ QWs [Fig. 7 (b)]: the hole concentrations in $In_{0.12}Ga_{0.88}N$ QWs then increase rapidly, resulting in more prominent spectrum 2 via recombination of e-h pairs in the $In_{0.12}Ga_{0.88}N$ QWs. It is pointed out that the carrier redistribution by the recombination dynamics is similar to the efficiency droop explained via the injection rate exceeding the recombination rate and subsequent carrier overflow or spill-over initiated by the saturation of the recombination rate (phase-space filling) [17],[20]. In our case, spectrum 2 via recombination of e-h pairs in the $In_{0.12}Ga_{0.88}N$ QWs is reinforced by redistribution of carriers with the injection rate exceeding the recombination rate in the $In_{0.15}Ga_{0.85}N$ QWs and subsequent spill-over to adjacent higher-bandgap QWs. Consequently, the limitation of recombination rate in lowest energy state in active region introduced by such as local potential fluctuation and quantum confined stark effect (QCSE) should be considered important in the onset of carrier spill-over and can be though to play an important role in terms of quantum efficiency and in InGaN-based LEDs [21]-[22].

## 4. Conclusion

In summary, we have investigated on the dominant factors determining the carrier distribution in InGaN/GaN MQW LED devices. A dichromatic MQW structure has been employed to understand how the carrier distribution is determined. By analyzing PL and temperature-dependent EL spectra, it has been demonstrated that the carrier recombination rate plays an important role in determining the carrier distribution in the MQWs.

Through these experiments and analysis, it is concluded that the simple differences in electron and hole characteristics, i.e., mobility and effective mass, are not sufficient to explain the observed spectra. The carrier dynamics including the carrier relaxation, build-up, and recombination affected by the saturation phenomena should be considered as an important factor in determining the carrier distribution in the InGaN MQW active region. Therefore, to achieve highly efficient InGaN LED devices via realization of a large effective active volume through uniform carrier distribution, more attention should be paid on how to increase the recombination rate in the MQWs, especially the radiative recombination rate, by using such measures as relaxing the piezoelectric field and/or local potential fluctuation.

**Figure Captions**

FIG. 1. Schematic diagram of (a) the LED device structure and (b) the MQW layer structure of the sample used in this study.

FIG. 2. Measurement results of the PL spectrum pumped by a 325-nm He-Cd laser, plotted on (a) linear and (b) semi-log scales. The dashed lines in (a) are Gaussian fitting curves.

FIG. 3. Measurement results of the EL spectrum as a function of driving current at 300 K. Each spectrum is normalized to its peak intensity.

FIG. 4. Measurement results of the EL spectrum from 50 to 200 K for injection currents of (a) 10 μA and (b) 10 mA. Each spectrum is normalized to its peak intensity.

FIG. 5. Measurement results for the emission intensity of spectrum 1 and spectrum 2 as a function of driving current at (a) 50 K and (b) 200 K.

FIG. 6. Numerical simulation results of the two-dimensional (a) electron, (b) hole, (c) radiative-recombination-rate distribution characteristics at low (3.5 A/cm$^2$) and high (70 A/cm$^2$) current densities. Each data is normalized to its peak value for comparison.

FIG. 7. Schematic energy bands diagram, including the carrier distribution and recombination in the MQW layer under (a) a low current density and (b) a high current density.

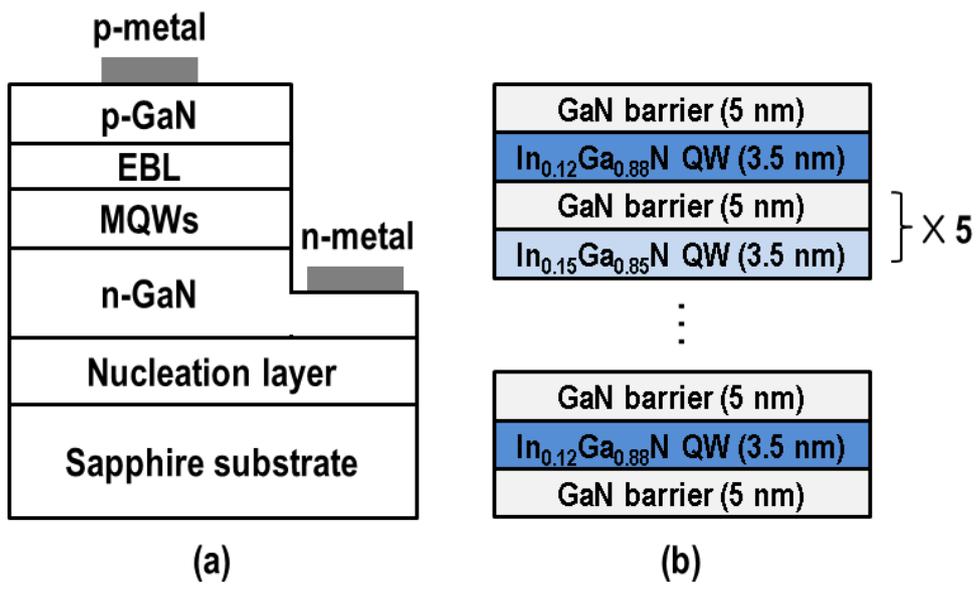

Figure 1.

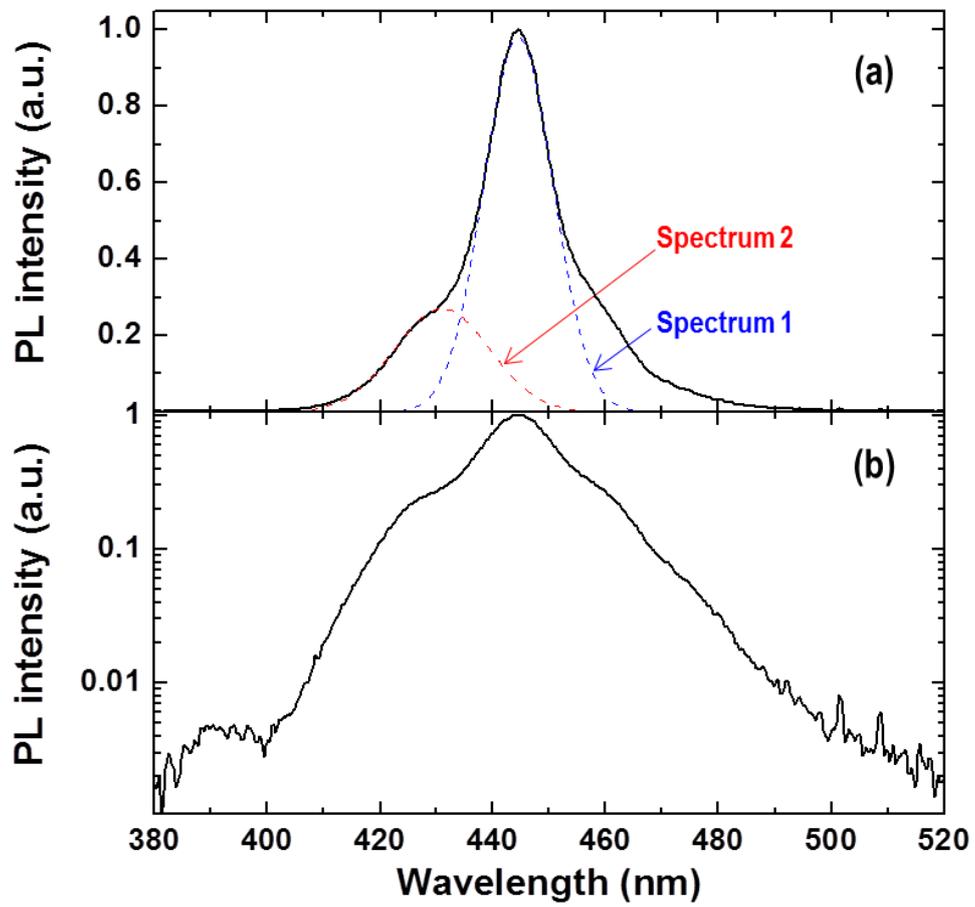

Figure 2.

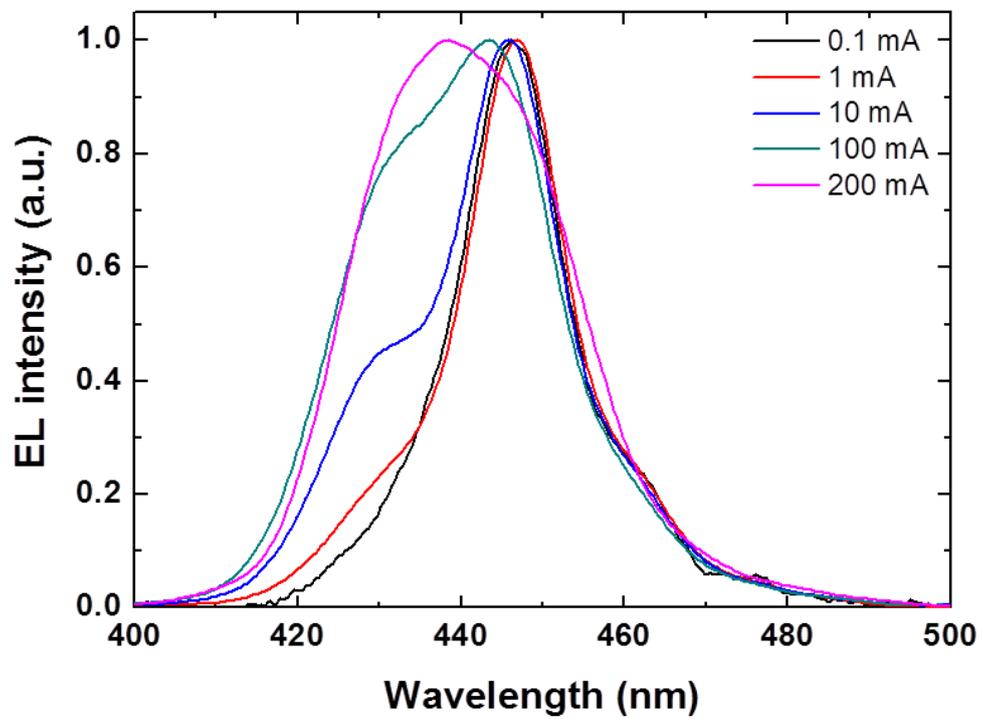

Figure 3.

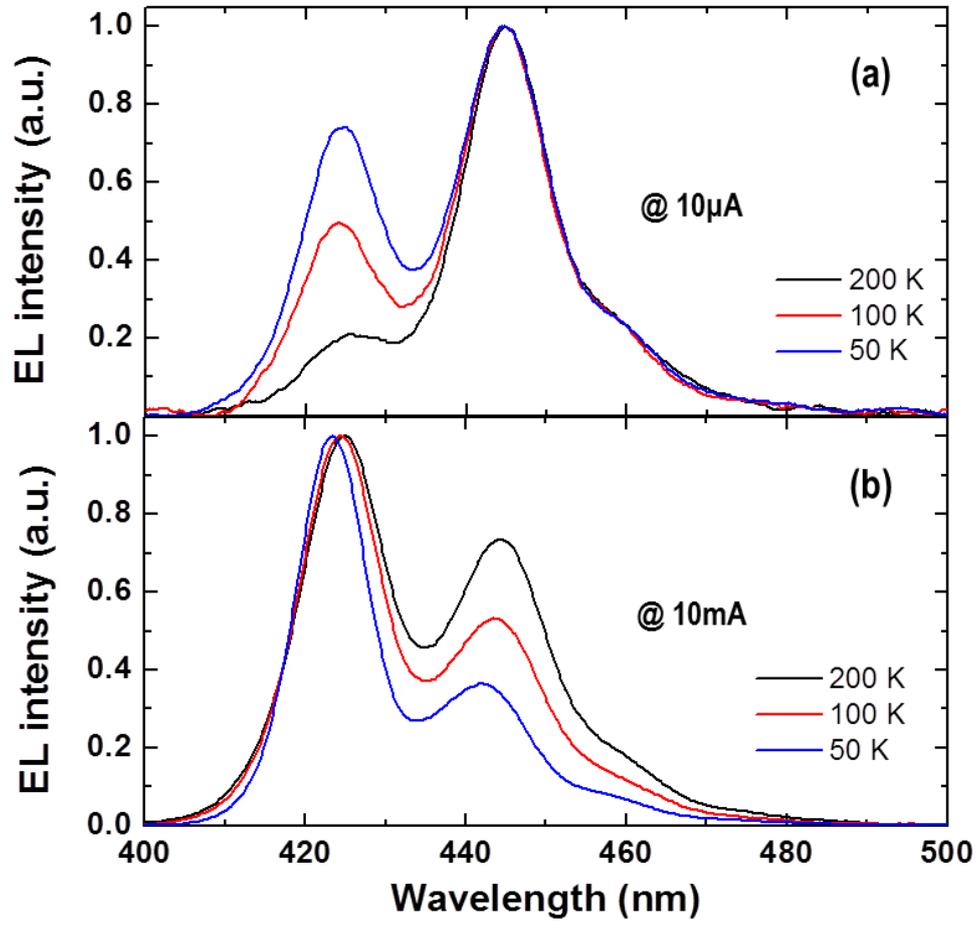

Figure 4.

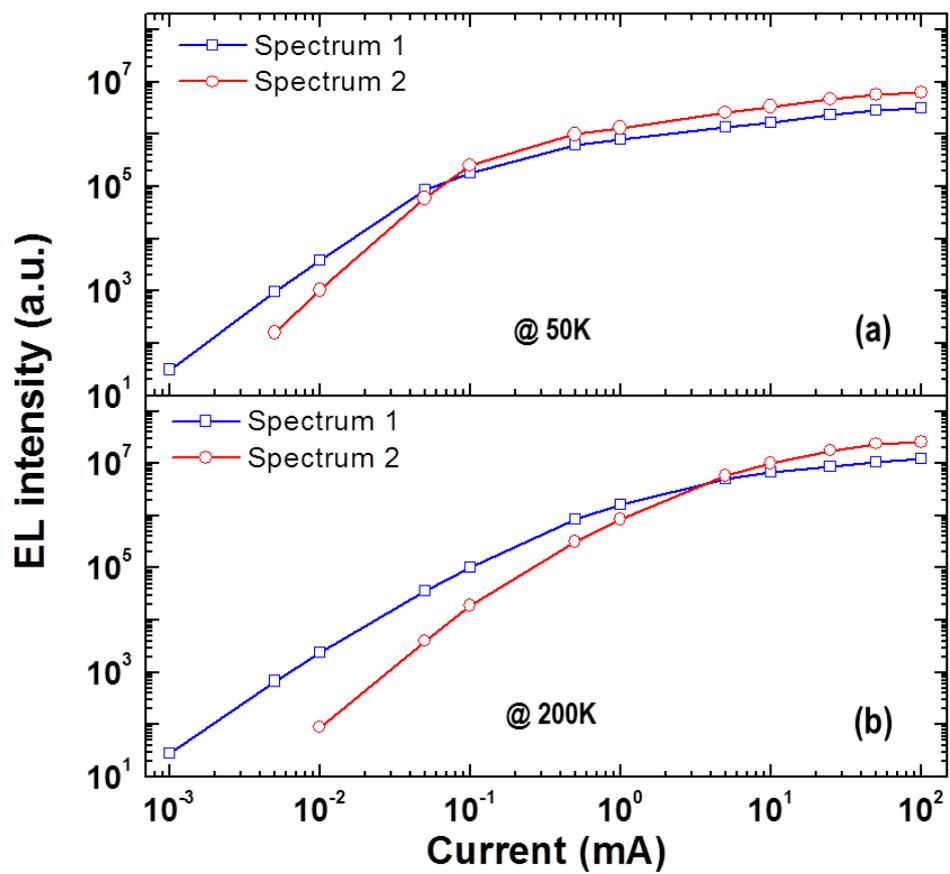

Figure 5.

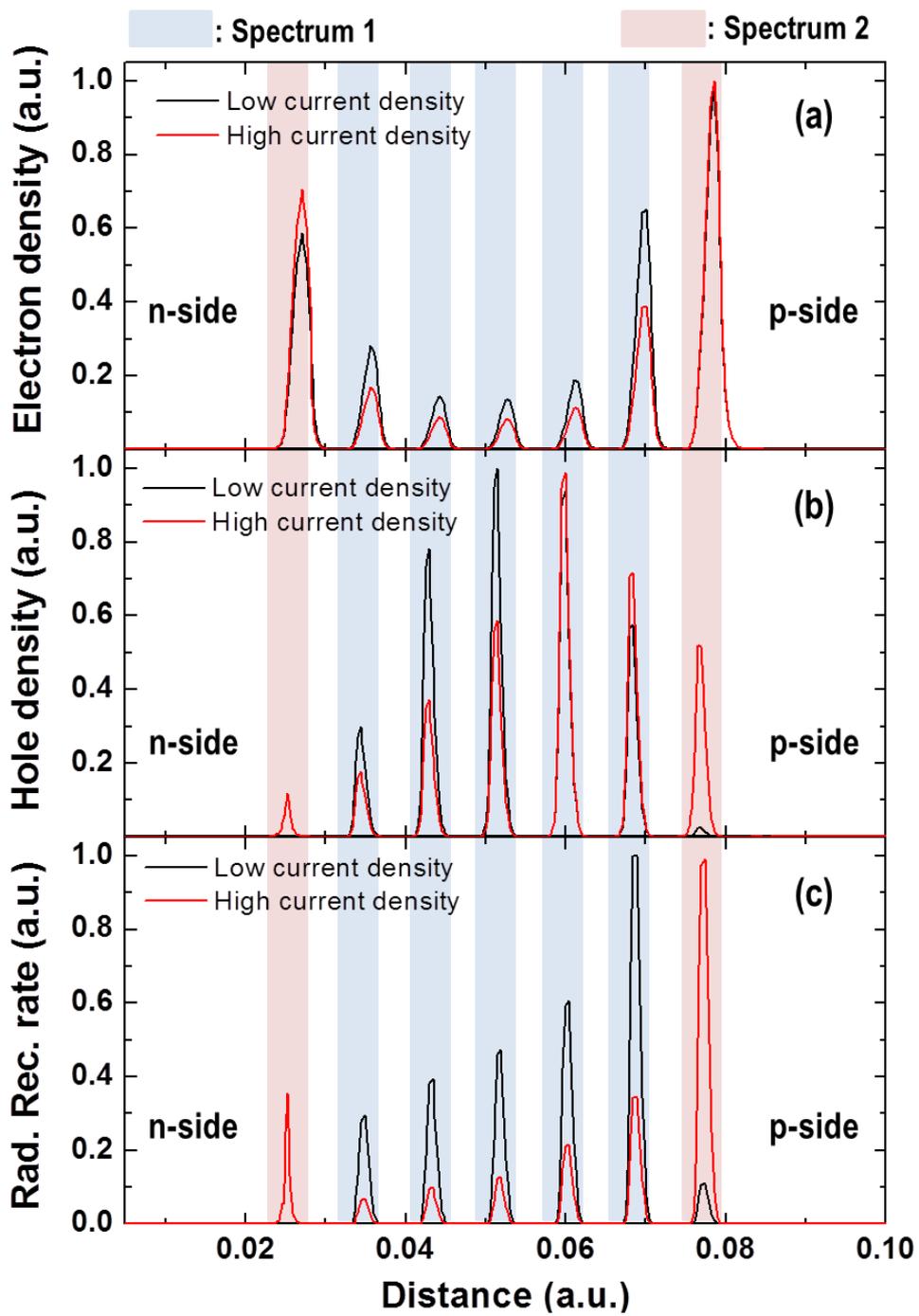

Figure 6.

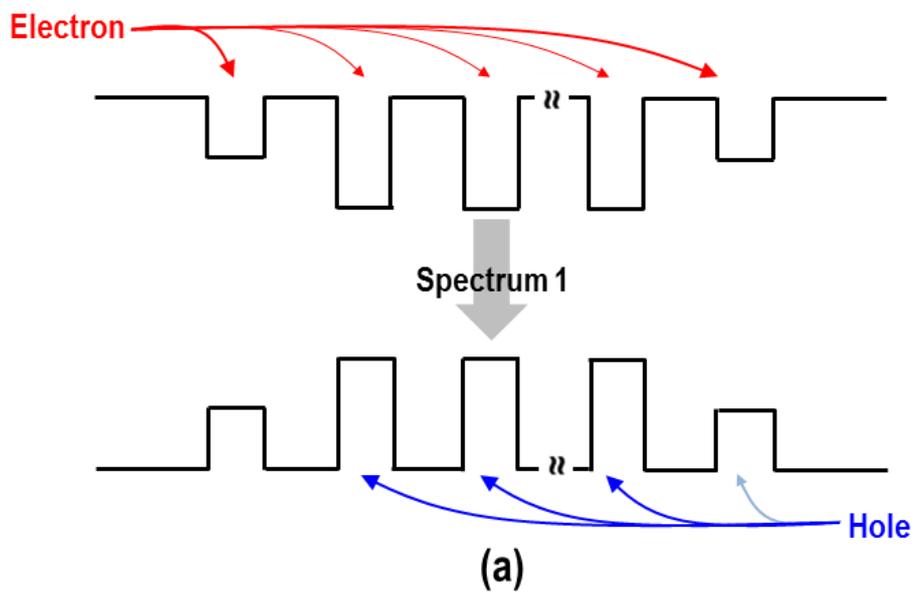

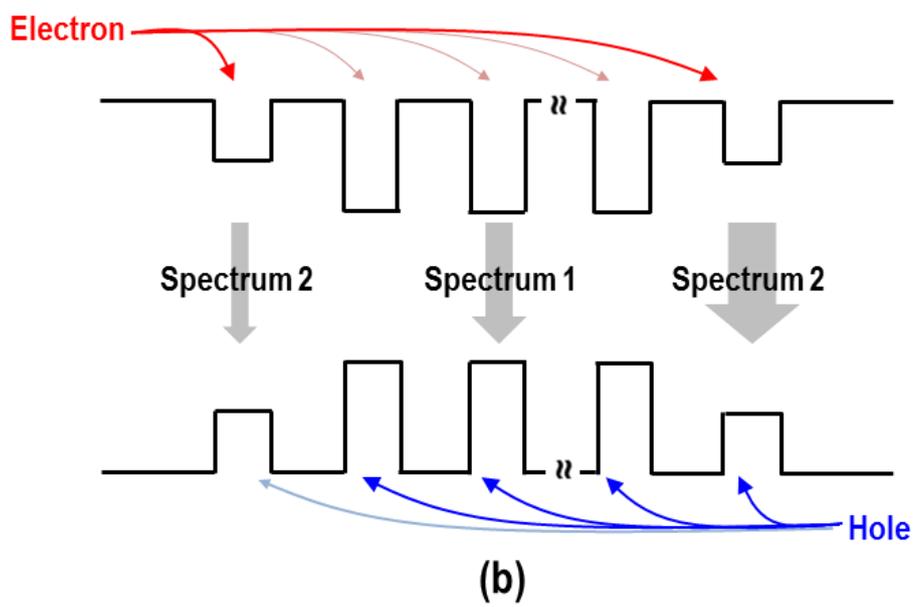

Figure 7.